\documentclass[a4paper]{spie}  

\usepackage[]{graphicx}
\usepackage{graphicx}
\usepackage{amssymb,amsmath,amsfonts}

\title{HD\,139614: the interferometric case for a group-Ib pre-transitional young disk}

\author{Lucas~Labadie\supit{a}, 
Alexis~Matter\supit{b},
Alexander Kreplin\supit{c},
Bruno Lopez\supit{d},
Sebastian Wolf\supit{e},
Gerd Weigelt\supit{c},
Steve~Ertel\supit{f},
Jean-Philippe~Berger\supit{f}
Jorg-Uwe~Pott\supit{g},
William~C.~Danchi\supit{h}
\skiplinehalf
\supit{a} I.\,Physikalisches Institut, Universit\"at zu K\"oln, Z\"ulpicher Str. 77, 50937 K\"oln, Germany \\
\supit{b} UJF-Grenoble~1/CNRS-INSU, Institut de Plan\'etologie et d'Astrophysique de Grenoble (IPAG) UMR 5274, Grenoble, F-38041, France\\
\supit{c} Max Planck Institut f\"ur Radioastronomie, Auf dem H\"ugel 69, 53121 Bonn, Germany\\
\supit{d} Laboratoire Lagrange, CNRS UMR 7293, UNS--Observatoire de la C\^ote d’Azur BP 4229, F-06304 Nice Cedex 4, France\\
\supit{e} Universit\"at zu Kiel, Institut f\"ur Theoretische Physik und Astrophysik, Leibnitzstr. 15, 24098 Kiel, Germany\\
\supit{f} European Southern Observatory, Munich, Germany \\
\supit{g} Max Planck Institut f\"ur Astronomie, K\"onigstuhl 17, D-69117 Heidelberg, Germany\\
\supit{h} NASA/GSFC, Greenbelt, MD 20771, USA
}

\authorinfo{Further author information: (Send correspondence to L.Labadie)\\L.Labadie: E-mail: labadie@ph1.uni-koeln.de, Telephone:\,+49 221 470 3493
}

\begin{document} 
  \maketitle 

\begin{abstract}
The Herbig Ae star HD\,139614 is a group-Ib object, which featureless SED indicates disk flaring and a possible pre-transitional evolutionary stage. We present mid- and near-IR interferometric results collected with MIDI, AMBER and PIONIER with the aim of constraining the spatial structure of the 0.1-10 AU disk region and assess its possible multi-component structure. A two-component disk model composed of an optically thin 2-AU wide inner disk and an outer temperature-gradient disk starting at 5.6 AU reproduces well the observations. This is an additional argument to the idea that group-I HAeBe inner disks could be already in the “disk-clearing” transient stage. HD\,139614 will become a prime target for mid-IR interferometric imaging with the second-generation instrument MATISSE of the VLTI.
\end{abstract}


\keywords{Astrophysics, YSO, Circumstellar Disks, Stellar Interferometry, Infrared, High Angular Resolution}

\section{Introduction}\label{section1}

Understanding the time evolution of circumstellar disks around low-mass and intermediate-mass stars is of the highest importance to improve our knowledge of (exo)planet formation. It is currently proposed that the processes transforming gas rich disks into mature debris disks with planetary systems involve grain growth and settling, photo-evaporation by stellar UV photons or disk clearing by forming and gas-accreting planets. However, the mechanisms by which planet embryos form in the disk and evolve towards massive gaseous Jupiters or small rocky planets is still not well understood. In order to progress in this direction it is important to detect and identify peculiar spatial structures in circumstellar disks such as rings, gaps, asymmetries that may reveal planetary signatures at a particular evolutionary stage. Coupled to the modeling of spectral energy distributions, infrared long baseline interferometric observations is a unique tool to probe and test disk configurations at various ages.\\
\\
In this paper we present the results of the observing campaign on the young star HD\,139614 and the subsequent modeling work based on VLTI interferometric data taken with MIDI, AMBER and PIONIER. With the low-resolution (R$\sim$30) mid-infrared data, the warm dust emission in the 1--10\,AU region of the circumstellar disk is probed, while the H and K near-infrared data are sensitive to the hotter dust closer to the star and down to the dust sublimation radius. Our modeling has been based in parallel on the implementation of geometrical disk models and radiative transfer calculations. For a detailed review on the physics of disks and their spatial structure, we direct the reader to topical literature\cite{Dullemond2010,Armitage2011}.\\
\\
In Sect.~\ref{section2} we present the context of our study and the object HD\,139614. In Sect.~\ref{section3} and~\ref{section4} the observations are presented together with the qualitative results. In Sect.~\ref{section5} the modeling work is introduced and its outcome is discussed in Sect.~\ref{section6}. 
\begin{figure*}[t]
\centering
\includegraphics[width=9cm]{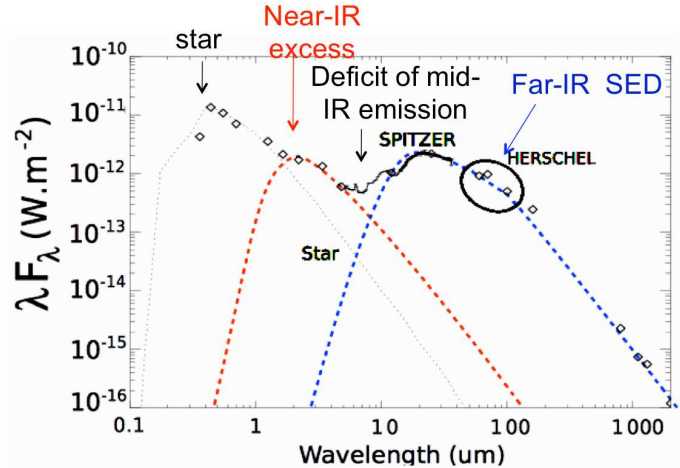}
\caption{Spectral energy distribution of HD\,139614 showing the near-infrared excess and the rising mid-infrared flux characteristic of pre-transitional disks. These two features can be modeled with simple hot ($T\gtrsim$1000\,K) and warm ($T\sim$200--400\,K) blackbody components.}\label{Fig0}
\end{figure*}

\section{HD\,139614 system in the context of group-I and group-II disks}\label{section2}

\begin{table*}[h]
\centering
\begin{tabular}{l c c c c c c c}
\hline\hline
$T_{\rm eff}$  & A$_V$       &    $M/M_\odot$     &  $\log$($L/L_\odot$)  & $R/R_\odot$   &  Age\,(Myr)           &  Sp.\,Typ  &  d\,(pc)    \\ \hline
 7600$\pm$300  &  0.5        &    1.76$\pm$0.2    &  1.10$\pm$0.15        & 2.06$\pm$0.4  &  8.8$^{+4.5}_{-1.9}$  &   A7V      &  142$\pm$27  \\ \hline
\end{tabular}
\caption{Stellar parameter of HD\,139614 taken from Alecian et al. (2013)\cite{Alecian2013}\,.}\label{table1}
\end{table*}
\noindent HD\,139614 is an intermediate-mass Herbig star with well characterized properties. In Table~\ref{table1}, we report the stellar parameters as derived in Alecian et al. (2013)\cite{Alecian2013}\,. It is a relatively evolved pre-main sequence star associated to the Lupus/Ophiuchus cloud. The spectral energy distribution (SED) exhibits a significant excess to the photospheric flux from the near-infrared to millimeter wavelengths. Its group-Ib “featureless” SED\cite{Meeus2001} suggests significant disk flaring as well as possible pre-transitional characteristics identified as a near-IR excess accompanied by a dip at 6 microns in the SED, then followed by a rising mid-infrared part (cf. Fig.~\ref{Fig0}).\\ 
In the Herbig HAeBe classification of Meeus et al. (2001)\cite{Meeus2001}\,, HD\,139614 is tagged as a group Ib object, meaning that its SED can be decomposed into a power law and a blackbody components and that it presents no silicates emission feature around 10\,$\mu$m. This class is different than the group II(b), in which the SED is described solely by a single power law. The index $a$ or $b$ refer to, respectively, the presence or absence of a significant 10\,$\mu$m silicate feature. \\
This classification is interesting from an evolutionary point of view in which a link between the two configurations is proposed: group I sources would host young hydrostatic disks with flaring geometry that produce the large mid- and far-infrared flux. Then these objects would evolve towards group II sources, which exhibit a flatter outer disk due to the settling of larger coagulated dust grains into the mid-plane. This results in a bluer and steeper infrared SED. The role of the puffed-up inner rim\cite{Dullemond2004} in shadowing part of the disk of group II sources is also advanced.  \\
Recent works at high angular resolution have evidenced the presence of large gaps in disks classified as group I objects\cite{Maaskant2013}\,. In such gapped disks, significant amount of mid- to far-IR emission in the SED can arise from large vertical walls at the inner edge of the outer disks that are directly irradiated by the central star. In addition, the presence or lack of silicate emission feature in the SED can also be interpreted in terms of gap size within the region where small and warm silicate grains ($\sim$300\,K) are normally efficiently emitting. In contrast, for group II objects with flatter disks no gap has been reported so far by spatially resolved observations, while none of the studied group II objects show any {\it absence} of the silicate emission feature. The difference in the spatial structure of the inner disk regions between group I and group II sources is still an open question, and it is beyond the goal of this paper to discuss in details the conditions of evolution of HAeBe's circumstellar disk. It is however interesting to note how recent spatially resolved observations have triggered new hypothesis on the possible evolutionary connection between group I and group II sources. It is suggested that these two classes could result from a bifurcation in the evolution of a common flared and gapless disk progenitor\cite{Maaskant2013}\,. As the formation of gaps in protoplanetary disks has notable implications on the process of planet formation, it is essential to gather additional knowledge on a larger sample of the so-called pre-transitional disks 

\section{Observations and data reduction}\label{section3}
\begin{figure*}[t]
\centering
\includegraphics[width=14.1cm]{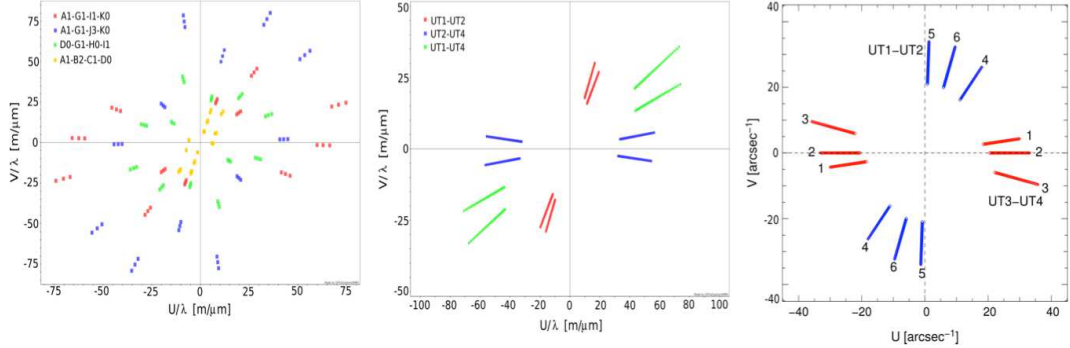}
\caption{uv coverage obtained for HD\,139614 with PIONIER (left panel), AMBER (center panel) and MIDI (right panel). Note that only the coverage of the quadruplet A1-G1-K0-I1 is displayed.}\label{Fig1}
\end{figure*}
HD\,139614 was observed in several campaigns with the VLTI from 2010 to 2013. The MIDI\cite{Leinert2003} observations were carried out with the 8-m unit telescopes in visitor mode on April 26$^{th}$ 2010 and in service mode on April 14$^{th}$ and 18$^{th}$ 2011. The observations were conducted using the popular HIGH-SENS mode of MIDI, which offers a resolution of R$\sim$30 at 10\,$\mu$m. We used the baselines UT3-UT4 and UT2-UT3 for these runs. In the near-infrared, we first conducted AMBER\cite{Petrov2007} observations in service mode on May 9$^{th}$ 2012 using the H+K mode in low spectral resolution. We used the UT1-UT3-UT4 triplet under good seeing conditions ($\sim$0.8$^{\prime\prime}$). Additional H band data have been acquired with PIONIER\cite{LeBouquin2011} using the 1.8-m auxiliary telescopes between March 2012 and July 2013 to improve the uv coverage for our object. PIONIER can recombine quadruplets instead of triplets as in the case of AMBER. In addition to the squared visibilities V$^2$, both AMBER and PIONIER deliver closure phases using the different available triplets. However, the closure phase measurements are not directly exploited in this paper. All the observations were obtained at low airmass ($\sim$1.1). In Table~\ref{table2} the logs of the different observations are presented together with the corresponding calibrator stars.
\begin{table*}[t]
\centering
\begin{tabular}{c c c c c}
\hline\hline
Date	    & UT    & Baseline   &  Calibrator   & Seeing($^{\prime\prime}$)    \\ \hline \hline
{} & {} & {MIDI} & {} & {} \\ 
{} & {} & {FOV of 300\,mas (40\,AU)} & {} & {} \\ \hline
26/04/2010  & 03:16:05     &    UT3-UT4     & HIP\,60979   & 1.2   \\ 
									& &     & HIP\,63066   &       \\  
26/04/2010  & 04:02:23     &    UT3-UT4     & HIP\,76552   & 1.1     \\ 
  &      &    						        & HIP\,76397   &       \\ 
26/04/2010  & 05:35:51     &    UT3-UT4     & HIP\,76397   & 1.3     \\ 
      					   &      &         & HIP\,76552   &       \\ 
14/04/2011  & 07:58:08     &    UT1-UT2     & HIP\,76397   & 1.4   \\ 
18/04/2011  & 03:35:05     &    UT1-UT2     & HIP\,72010   & 0.7   \\ 
 &     							&           & HIP\,74395   &    \\ 
18/04/2011  & 05:15:58     &    UT1-UT2     & HIP\,76552   & 1.3   \\ \hline \hline
{} & {} & {AMBER} & {} & {} \\ 
{} & {} & {FOV of 60\,mas (9\,AU)} & {} & {} \\ \hline
09/05/2012  & 05:27:00     &    UT1-UT3-UT4     & HD\,141702   & 1.01     \\ 
09/05/2012  & 06:27:00     &    UT1-UT3-UT4     & HD\,140785   & 0.65     \\ \hline\hline
{} & {} & {PIONIER} & {} & {} \\ 
{} & {} & {FOV of 200\,mas (25\,AU)} & {} & {} \\ \hline
25/03/2012  & 08:55:41     &    A1-G1-K0-I1     &        & 1.3     \\ 
06/06/2013  & 01:27:00     &    A1-G1-J3-K0     &        & 1.1     \\ 
16/06/2013  & 23:31:00     &    D0-G1-H0-I1     &        & 1.0     \\ 
03/07/2013  & 23:58:00     &    A1-B2-C1-D0     &        & 1.1     \\ \hline\hline
\end{tabular}
\caption{Log of the observations of HD\,139614 with MIDI, AMBER and PIONIER.}\label{table2}
\end{table*}
The data reduction was achieved using the different pipelines publicly available such as EWS/MIA\cite{Jaffe2004} for MIDI or the amdlib package release 3.0.5 provided by the JMMC\footnote{www.jmmc.fr}. The MIDI calibrated visibilities and correlated fluxes span the 8--13\,$\mu$m spectral range, while the low spectral resolution AMBER squared visibilities span the 2.06--2.46\,$\mu$m range in 13 spectral channels. With PIONIER we obtain three visibilities per baseline in three spectral channel within the H band. Both mid- and near-infrared observations present a good enough uv coverage to constrain diameters at different position angles across HD\,139614's disk (see Fig.~\ref{Fig1}). \\
For purpose of modeling, we also compiled the SED of our object by retrieving optical photometry from the Tycho-2 catalog, near-infrared photometry from 2-MASS\cite{Skrutskie2006}\,, the SPITZER/IRS 5--40\,$\mu$m spectrum, the Herschel/PACS photometry at 70, 100 and 160\,$\mu$m, and millimeter photometry at 0.8\,mm, 1.1\,mm, 1.3\,mm and 2\,mm from Sylvester et al. (1997)\cite{Sylvester1997}\,. 

\section{Observational Results}\label{section4}

\begin{figure*}[b]
\centering
\includegraphics[width=13cm]{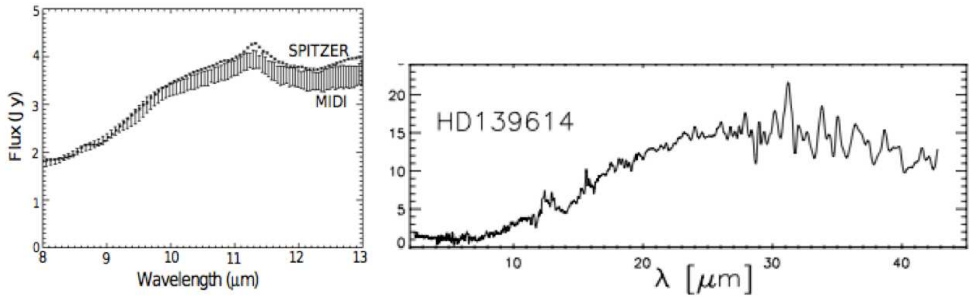}
\caption{Comparison between the MIDI, Spitzer and ISO spectra\cite{Meeus2001} of HD\,139614. The MIDI spectrum agrees well with Spitzer's one and exhibit the PAH emission feature at 11.3\,$\mu$m.}\label{Fig2}
\end{figure*}

\subsection{Spectral Energy Distribution}

The SED of HD\,139614 is essentially featureless and presents no clear indication of silicate emission around 10\,$\mu$m as inferred from the MIDI and the SPITZER/IRS spectra (see Fig.~\ref{Fig2}). The ISO spectrum published by Meeus et al. (2001) is displayed for comparison. The calibrated MIDI spectrum is a by-product of the EWS data reduction, and the error on the flux measurement was reduced by averaging the six flux measurements obtained for each uv point. The MIDI and Spitzer spectra agree very well over the whole N band, which suggest that no significant flux variability is detected. The PAH emission features seen the Spitzer spectrum at 8.6\,$\mu$m and 11.3\,$\mu$m are detected in the MIDI spectrum as well, although with less accuracy. In particular, the 11.3\,$\mu$m peak is clearly visible in the MIDI spectrum (see Sect.~\ref{section6}).

\subsection{Visibilities}

\begin{figure*}[h]
\centering
\includegraphics[width=7cm]{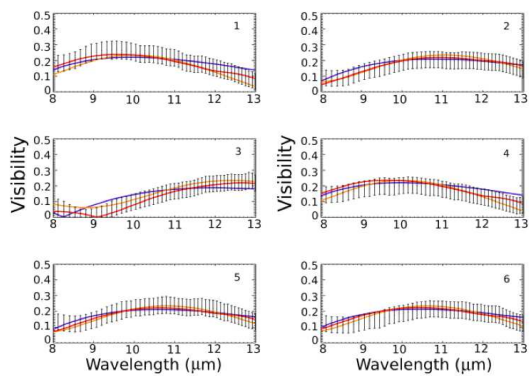}
\caption{Calibrated MIDI visibilities obtained with the different baselines shown in Fig.~\ref{Fig1} (in black with error bars). The uniform achromatic ring model is plotted in orange, the one-component disk model in blue, and the two-component disk model in red for the high-density case (see Sect.~\ref{MIDI}).}\label{Fig3}
\end{figure*}

\begin{figure*}[t]
\centering
\includegraphics[width=15cm]{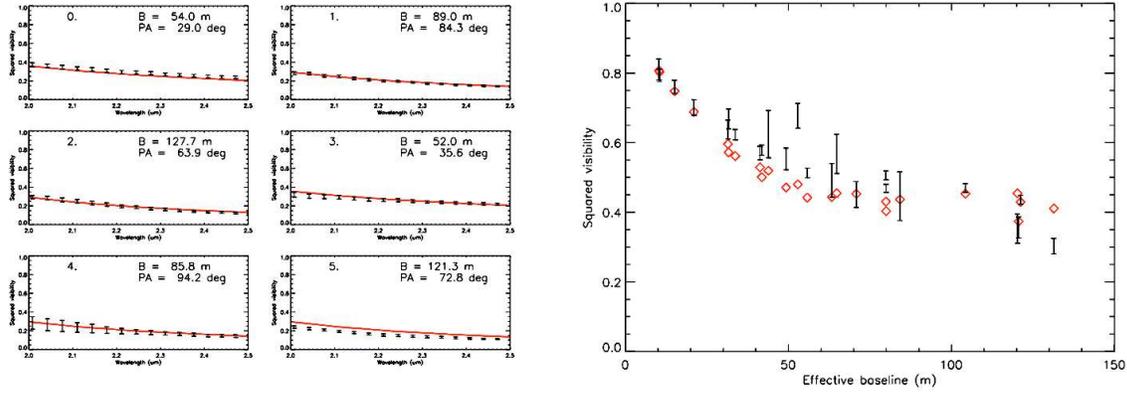}
\caption{AMBER calibrated squared visibilities as a function of the wavelength for the six UV points (left) and PIONIER visibilities as a function of baseline at 1.6\,$\mu$m (right). The red curves and data points correspond to the results of the radiative transfer modeling (see Sect.~\ref{AMBER}.)}\label{Fig4}
\end{figure*}

The spatial resolution achieved with our MIDI mid-infrared interferometric observations corresponds to $\sim$20\,mas ($\sim$3\,AU) at 10\,$\mu$m. In the near-IR, the observations are obtained both with the UTs (AMBER) and the ATs (PIONIER) baselines, spanning resolutions from $\sim$2 to 11\,mas (0.3--1.6\,AU). The calibrated visibility curves of MIDI shown in Fig.~\ref{Fig3} indicate that the source is fully resolved for all the uv points with visibilities varying between 0.05 and 0.2 that show sinusoidal variations. As the hypothesis of a resolved close-binary system is discarded\cite{Matter2014} such low visibilities indicate a significantly extended warm emitting region around HD\,139614.\\
\\
\noindent Furthermore the curves do not present the characteristic drop-off between 8 and 9\,$\mu$m as seen for several Herbig stars disks in Leinert et al. (2004)\cite{Leinert2004} and modeled by Van Boekel et al. (2005)\cite{vanBoekel2005} as the mid-infrared tail of the confined puffed-up inner rim, which implies high visibilities at 8\,$\mu$m followed by a sharp drop down to 9.5\,$\mu$m. The emission at 8\,$\mu$m appears quite resolved by MIDI, suggesting that if hot material heated by the star is present it must be relatively extended. In the near-infrared, the UT baselines mainly probe the high spatial frequencies and the hot dust in the system. As it can be seen in Fig.~\ref{Fig4}, the AMBER squared visibilities are quite low as well, i.e. varying between $\sim$0.1 and 0.4, which suggest hot material extending well beyond the 0.3\,AU distance theoretically resolved by the instrument. The PIONIER data also show resolved emission even with the shorter baselines ($\lesssim$40\,m) of the AT array, confirming that the hot dust emission is probably spatially extended. Although not shown in this article, both the MIDI differential phases and the AMBER and PIONIER closure phases did not reveal any clear spatial asymmetry in the spatial structure of the disk.

\section{Modeling}\label{section5}

\subsection{MIDI and spectral energy distribution data}\label{MIDI}

\begin{figure*}[t]
\centering
\includegraphics[width=8cm]{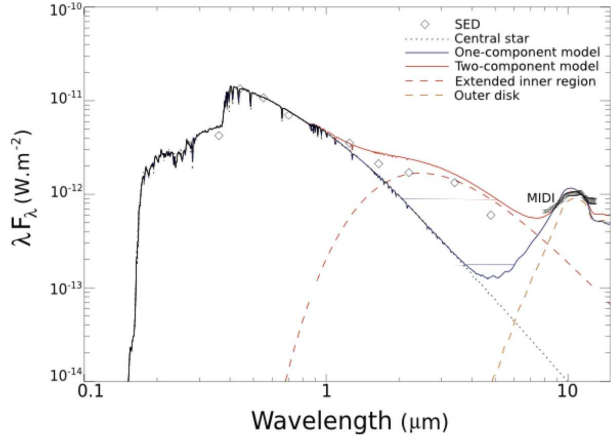}
\caption{Near- and mid-infrared SED of HD\,139614 and corresponding fit for the one-component and two-component models presented in this study for the high-density case}\label{Fig5}
\end{figure*}

The adopted methodology consisted in modeling the structure of the circumstellar disk using the spectral energy distribution and the VLTI interferometric visibilities. This is a commonly adopted approach that has demonstrated to be successful. We proceeded gradually from the modeling of the large scale parameters of the disk (inclination, P.A. etc...) using geometrical flat models to the more sophisticated 3D modeling of the inner disk using the RADMC-3D radiative transfer code\cite{Dullemond2012}\,. 
As a first step, we constrained the spatial extent of the 10\,$\mu$m warm emitting region using solely the mid-IR visibilities. We applied a temperature-independent uniform ring model parametrized by the inner radius of the ring $R_{\rm in}$, its width $\Delta R$, inclination $i$ and position angle $\theta$, and the photospheric contribution $f_{\rm in}$ to the total flux\cite{Fedele2008}\,. The $\chi^2$ best fit obtained with two different grid resolutions\cite{Matter2014} suggests that the bulk of the mid-infrared emission originates from a region extending from 4.5\,AU$\pm$0.2\,AU to 8\,AU$\pm$0.2\,AU. A low inclination of the ring of $i$=20$^{\circ}$$\pm$2.1$^{\circ}$ is derived, in agreement with previously published measurements\cite{Meeus1998,Panic2009}\,. A position angle of $\theta$=112$^{\circ}$$\pm$9.3$^{\circ}$ is found. \\
We then adopted a more physical temperature-gradient disk model characterized by the temperature and surface density distributions $T_r$\,=\,$T_{in}(r/r_{in})^{-q}$ and $\Sigma_r$\,=\,$\Sigma_{in}(r/r_{in})^{-p}$. The first scenario implied a one-component disk model (i.e. central star+temperature-gradient disk), which parameters were optimized to fit simultaneously the MIDI visibilities and the near- and mid-infrared spectral energy distribution. The stellar SED was modeled with a Kurucz synthetic spectrum with $T_{\rm eff}$=7750\,K. A number of parameters were fixed in advance. These are the power law exponent of the surface density that was assumed to be $p$=3/2, a pure iron-free olivine composition\cite{Juhasz2010} and the outer radius fixed to $R_{\rm out}$=100\,AU. For the disk surface density $\Sigma_{\rm in}$ at $r_{\rm in}$ we assumed three different cases of low (2$\times$10$^{-5}$g.cm$^{-2}$), medium (6$\times$10$^{-5}$g.cm$^{-2}$) and high (10$^{-4}$g.cm$^{-2}$) density to span a range of optical depth values (0.01\,$\leq$$\tau$$\leq$\,0.1) that allows the overall shape of the mid-infrared spectrum to be reproduced. We assume $i$=20$^{\circ}$ and $\theta$=112$^{\circ}$ from the previous analysis. The free parameters for the one-component model are thus the inner radius $r_{\rm in}$, the temperature $T_{\rm in}$ at $r_{\rm in}$, and the temperature power law exponent $q$. A simultaneous modeling of the SED and MIDI visibilities with $\chi^2$ minimization resulted in the models shown in Table~\ref{table3}. 
\begin{table*}[b]
\centering
\begin{tabular}{l c c c c c c c}
\hline\hline
Model	  				&  Surface density  & $\Delta r_{\rm hot}$\,(AU) & $\log(\epsilon)$ &   $r_{\rm in}$\,(AU)      &   $T_{\rm in}$\,K   &  q   &  $\chi^{2}_r$        \\ \hline
{}		  	       		&	low 	 	    & 				--			 &				    &  2.79$\pm$0.06 			&  669$\pm$24	      & 0.74$\pm$0.06   &  3.2  	\\ 
One-component disk		&	medium	    	& 			 	-- 			 & 		    		&  2.78$\pm$0.09 			&  478$\pm$12	      & 0.66$\pm$0.06   &  3.0		\\ 
{}		  				&	high	    	& 				-- 			 &		    		&  2.77$\pm$0.09 			&  421$\pm$9		  & 0.62$\pm$0.06   &  3.4		\\ \hline
{}		  				&	low 	    	& 		 1.66$\pm$0.12		 &	-2.78$\pm$0.1   &  5.60$\pm$0.10 			&  445$\pm$33	      & 0.78$\pm$0.18   &  1.6  	\\ 
Two-component disk		&	medium	    	& 		 2.04$\pm$0.21	   	 &	-2.87$\pm$0.12  &  5.61$\pm$0.12 			&  365$\pm$19	      & 0.97$\pm$0.21   &  1.04		\\ 
{}		  				&	high	    	& 		 2.13$\pm$0.12		 &  -2.87$\pm$0.06  &  5.63$\pm$0.12 			&  332$\pm$9		  & 1.04$\pm$0.06   &  0.95		\\ \hline\hline

\end{tabular}
\caption{Results of the disk modeling after $\chi^2$ minimization based on the MIDI visibilities and near+mid-infrared spectral energy distribution with associated 3-$\sigma$ error bars.}\label{table3}
\end{table*}
We see that while the position of the inner rim of the outer disk is well constrained around 2.8\,AU, the corresponding temperatures are quite scattered and high for a distance of $\sim$3\,AU. Note that for the binary star V892\,Tau, Monnier et al. (2008)\cite{Monnier2008} derived a temperature of $T\sim$450\,K at the inner rim of the circumbinary disk, which was found to be as far as $\sim$17\,AU from the central star. However the modeling of the authors assumed the equivalent luminosity of 400\,$L_\odot$ for two B8V stars. In all three cases, the best-fit models fail to reproduce both the SED and visibilities, in particular because almost no near-infrared excess is predicted. Smaller values for $r_{\rm in}$ obviously increases the near-infrared contribution in the model but also strongly disrupts the visibilities fit by predicting high values at 8\,$\mu$m resulting from unresolved hot material close to the star. As the uniform ring model already predicted, the outer disk inner rim could be located further away, between 4.5\,AU and 8\,AU. This suggests a more sophisticated spatial structure, with for instance the presence of a hot inner component responsible for the near-infrared excess. We hence tested a second temperature-gradient model by introducing an inner component spatially separated from the outer disk. Such a structure is found in gapped pre-transitional disks\cite{Espaillat2008,Espaillat2010}\,.\\
\\
For our new two-component temperature gradient model, we assumed a close-in hot component with a constant temperature of $T\sim$1500\,K that produces the 3--4\,$\mu$m excess typically observed for Herbig stars\cite{Natta2001}\,. The hot dust component is assumed to be as close as $r$=0.22\,AU, the dust sublimation radius derived from the luminosity of HD\,139614. In addition to the parameter already defined for the one-component model, we introduce two additional free parameters, namely the radial width of the hot dust $\Delta r_{\rm hot}$ and the emissivity factor $\epsilon$ to include the effect of the dust optical depth. 
The results of the $\chi^2$ minimization are displayed in Table~\ref{table3}. In addition to an improved result for the reduced $\chi^2$, the agreement with the observed SED near-infrared is much improved (see Fig.~\ref{Fig5}). The fit with the MIDI visibilities is still excellent as it can be seen in Fig.~\ref{Fig3}. 
Our best solution based on the combination of the MIDI visibilities and SED data has two components: an optically thin ($\epsilon\sim$10$^{-2.87}$) inner component extending from the dust sublimation radius to about 2.3\,AU; a narrower hot component produces visibilities at 8\,$\mu$m that significantly diverge from the observations; an optically thick outer disk starting at  $\sim$5.6\,AU with a temperature of 332 K. However, as in the one-component disk case, our temperature estimate still exceeds the upper limit temperature (≈270 K) expected at 5.6 AU, under the assumption of a pure silicate dust composition.

\subsection{Near-infrared observations with AMBER and PIONIER}\label{AMBER}

As the possible hot inner component predominantly contributes to the near-infrared flux, the MIDI observations are not ideally suited to constrain unambiguously the dust structure within 1--2\,AU from the star. The AMBER and PIONIER observations in the K and H bands offer the opportunity to confirm and refine our previous results. Therefore, we have implemented a self-consistent radiative transfer modeling of the spatial structure of the disk components. 
The code RADMC-3D code was used in this work.  Based on a Monte Carlo approach, the code computes the radiation field and the disk temperature distribution from a given dust density distribution, and provides SEDs and synthetic images at different wavelengths. 
The dust surface density $\Sigma(r)$ and scale height $H(r)$ are parameterized as $\Sigma(r)$\,=\,$\Sigma_{\rm out}(r/r_{out})^{p}$ and $H(r)$\,=\,$H_{\rm out}(r/r_{out})^{\beta}$, where $\Sigma_{\rm out}$ and $H_{\rm out}$ are the surface density and scale height at the disk outer radius $r_{\rm out}$, p and $\beta$ the power-law exponent and flaring index.\\
The properties of the outer disk (power law exponent $p$, disk mass of the outer component $M_{\rm out}$\footnote{Total gas+dust, assuming gas-to-dust mass ratio of 100.}, inner radius of the outer disk $R_{\rm in}$, dust composition) are constrained by the far-IR and sub-mm SED data points.
\begin{figure*}[t]
\centering
\includegraphics[width=10cm]{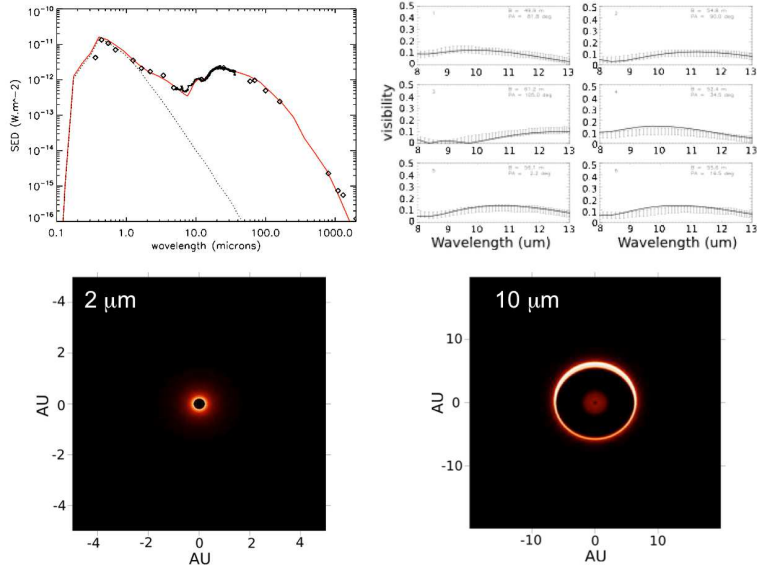}
\caption{{\it Top left:} synthetic SED of HD\,139614 obtained by radiative transfer modeling with the parameters shown in Table~\ref{tab5}. {\it Top right:} fit to the experimental mid-IR visibilities. {\it Bottom:} Images of the disk of HD\,139614 as seen at 2\,$\mu$m and 10\,$\mu$m with the star removed. We adopted a linear scale. The inclination $i$=20$^{\circ}$ is taken into account here.}\label{Fig6}
\end{figure*}\\
Then assuming co-planarity with the outer disk, the radiative transfer modeling of the inner disk is performed using a $\chi^2$ minimization exploring a 10$\times$10$\times$10 grid on the three free parameters $r_{\rm out}$, inner disk total mass $M_{\rm hot}$\footnote{With the same assumptions as for the outer disk} and surface density index $p$. We set the inner radius to the expected sublimation radius of HD\,139614, namely $\sim$0.2\,AU. The minimization is bounded by different values of fixed parameters as the grain size distribution, the dust composition, and the ratio $(H/r)_{\rm out}$. The accurate position of the outer disk’s inner radius $R_{\rm in}$ and the corresponding scale height is determined in a third step based on the MIDI visibilities. The numerical results of our modeling are summarized in the Table~\ref{tab5}.
\begin{table*}[h]
\centering
\begin{tabular}{c c c c c c }
\hline\hline
{}                				 & {} 	 			    &   {}				   	    	& {Inner hot component} 			& {} 			               & {}				    \\ \hline
$r_{\rm out}$ (AU) 	 		     &  $p$   				&   $M_{\rm hot}$/$M_{\odot}$   &  grain size dist. ($\mu$m) & composition                  &   $(H/r)_{\rm out}$\\ \hline
2.57 ($r_{\rm sub}$=0.22\,AU)    &  0.64 ($\beta$=2/7)  &   8.8$\times$10$^{-9}$  		&  5--20		                    & 80\% olivine,  &  0.1 at 2.6\,AU    \\ 
   								 &   		   		    & 		  				  		&        		                    & 20\% graphite  &      \\ \hline \hline
{}              		         & {} 	 			    & {} 						    & {Outer disk} 			            & {} 				           & {}				     \\ \hline
$R_{\rm in}$ (AU) 	 			 &  $p$   				&   $M_{\rm out}$/$M_{\odot}$  	&  grain size dist. ($\mu$m) & composition                  &   $(H/r)_{\rm out}$  \\ \hline
6.3 ($R_{\rm out}$=150\,AU)	 	 &  -1 ($\beta$=1/7)    &   1.3$\times$10$^{-2}$        &  0.1--3000		                & 100\% olivine                &  0.14 at 150\,AU     \\ \hline \hline
\end{tabular}
\caption{Results of the radiative transfer modeling of the multi-component disk of HD\,139614.}\label{tab5}
\end{table*}\\
In Fig.~\ref{Fig6} we show the results of the radiative transfer modeling and compare it to the observational data. The synthetic spectral energy distribution matches very well with the photometry from the optical to the millimeter range. The interferometric data are also quite well reproduced in the mid-infrared (cf. Fig.~\ref{Fig6}, upper right), and in the near-infrared (cf. Fig.~\ref{Fig4} with the AMBER and PIONIER data). Synthetic images at 2\,$\mu$m and 10\,$\mu$m of the disk around HD\,139614 have been produced using RADMC-3D as shown in Fig.~\ref{Fig6}, where they are plotted in a linear scale and with the central star removed. Not seen in these images is the tenuous dust that may still survive in the gap and which is needed to reproduce the excess emission in the SED around 6\,$\mu$m. An upper limit of $M_{\rm dust,gap}$$\leq$10$^{-10}$\,$M_{\odot}$ is derived, above which the modeled interferometric visibilities clearly diverge from the measured quantities.

\section{Discussion and conclusions}\label{section6}

We show in this paper that the combination of near-infrared and mid-infrared interferometric data can strongly constrain the spatial structure of the disk around HD\,139614. In particular, the interferometric data have ruled out with good confidence the hypothesis of a continuous disk in hydrostatic equilibrium and inwardly truncated at the dust sublimation radius. This reaffirms the power of long-baseline interferometry as a prime tool for studying young disks and their evolutionary trends.
\begin{itemize}
\item HD\,139614 appears to host an additional example of pre-transitional disks, i.e. a gapped disk found around both T Tauri and Herbig stars that could originate from the clearing process triggered by a young protoplanet. The possibility that the gap is in reality a strongly shadowed area of the disk is remote, as this  would imply a very large scale height of the near-infrared disk with a brightness impacting strongly the near-infrared visibilities and at the same time a substantial amount of cold dust radiating at millimeter wavelength.
\item Our simulations indicate that instead of a confined optically thick and narrow ring of dust, the near-infrared excess is better modeled by a extended and optically thin dust component of $\sim$2.5\,AU width with an increasing dust density outward to the disk (positive parameter $p$) in order to match the level of mid-IR visibilities around 8\,$\mu$m. The near-infrared visibility values do not contradict this hypothesis. 
\item Preliminary results indicate that the scale height of the outer disk at its inner edge plays a significant role in reprocessing the mid-infrared flux and that its temperature should be relatively high ($\sim$350\,K). Further work will better constrain this quantity.     
\item While not discuss in this paper, the PAH emission feature at 11.3\,$\mu$m is detected in the MIDI uncorrelated spectrum. Although the level of uncertainty on the MIDI visibilities does not permit to draw firm conclusions, we have indications on the fact that the PAH emitting region should be comparable in size, or even a bit larger, than the mid-IR continuum emission region.
\item None of near- and mid-infrared interferometric closure phases exhibit noticeable signatures. It is very unlikely that HD\,139614 could be in reality a tight binary system with a low-mass infrared companion.
\end{itemize}
\begin{table*}[b]
\centering
\begin{tabular}{l c c c c c c }
\hline\hline
{}             							  & Age (Myr)     &  Spectral type	   	  & $r_{\rm i}$ (AU)$^{1}$ 	  & $r_{\rm out}$ (AU)      & D (pc) 	&	 Gap size (AU)	    \\ \hline
HD\,139614\cite{Matter2014}				  & 8 		      &  A7		   		 	  & 0.2 	 			      & 2.6                     & 	140     &    3.7	    \\ 
HD\,100546\cite{Benisty2010,Tatulli2011}  & 10 		      &  B9		   		 	  & 0.26 	 			      & $\sim$4                 & 	103     &    9	     \\ 
HD\,169142\cite{Maaskant2013}		   	  & 13.5 	      &  A5V		   		  & $\sim$0.1 	 		      & 0.2\,(?)    	        & 	145     &    $<$23\,(?)	     \\ 
IRS\,48\cite{Maaskant2013}				  & 15   	      &  A0		   		 	  & $\sim$0.1 	 		      & 0.3		                & 	120     &    $<$63	     \\ 
HD\,97048\cite{Maaskant2013} 			  & --	     	  &  B9.5		   	 	  & 0.3		   			      & 2.5		                & 	158     &    $<$34	        \\ \hline
HD\,135344\,B\cite{Maaskant2013}		  & 6	     	  &  F4V		   		  & $\sim$0.1 	 		      & 0.3\,(?)		        & 	140     &    $<$30\,(?)	     \\ 
T\,Cha\cite{Olofsson2013}				  & 7	    	  &  K0-G8		   	 	  & 0.07 	 		      	  & 0.11		            & 	100     &    12	     \\ 
LkCa\,15\cite{Espaillat2010}			  & 3-5	  	      &  K5-K3		   	 	  & 0.015 	 		      	  & 0.19		            & 	140     &    58	     \\ 
Rox\,44\cite{Espaillat2010}   			  &  		  	  &  K3		   		 	  & 0.25 	 		      	  & 0.4		                & 	120     &    36	     \\ 
UX\,Tau\cite{Espaillat2010}	  			  &  1		      &  G8		   		 	  & 0.15 	 		          & 0.4		                & 	120     &    36	     \\ \hline
\end{tabular}
\caption{Preliminary compilation of some known pre-transitional objects already published. The ages are given here without error bars, but an uncertainty of about several Myr should be accounted for. Hence we only have a trend on the age of these objects the ones respect to the others. (1): in most cased $r_{\rm i}$ is identified with the dust sublimation radius.}\label{tab6}
\end{table*}
\noindent Our high-resolution study of HD139614, a star classified as a group-Ib HAeBe source, strongly suggests the pre-transitional nature of its circumstellar disk, with a low dust-density gap extending over few astronomical units, and a relatively spatially extended and optically thin hot inner component. It is interesting to place this result in the larger context of HAeBe and T-Tauri pre-transitional objects: one finds for instance that the gaps in HD139614 and HD100546 are surprisingly narrow in comparison to the T-Tauri disk gaps. In the latter objects however, the inner hot component was determined to be spatially compact. A non-exhaustive compilation of currently known pre-transitional disks is presented in Table~\ref{tab6}. Note that the ages are indicative and show, accordingly to the literature, uncertainties of several Myrs.\\
Because of the importance of gapped disked in the understanding of planet formation, our result brings additional insight into this debate and propose an additional candidate to better understand the origin of disk dispersal mechanisms in different classes of objects. HD\,139614 will become a prime target for mid-IR interferometric imaging with the second-generation instrument MATISSE of the VLTI.



\bibliography{report}   
\bibliographystyle{spiebib}   

\end{document}